\documentclass[aps,prb,citeautoscript,superscriptaddress,amsmath,amssymb]{revtex4-2} 
\usepackage{tikz} 
\usepackage{graphicx}
\usepackage{dcolumn}
\usepackage{lettrine}
\usepackage{soul}
\usepackage{bm}
\usepackage{hyperref}
 \usepackage{float} 
\usepackage{booktabs}
\usepackage[version=4]{mhchem}
\usepackage[normalem]{ulem}
\usepackage{ragged2e} 
\usepackage{comment}
\usepackage[T1]{fontenc}
\usepackage[utf8]{inputenc}

\newcommand{\SL}[2]{[#1\!\parallel\!#2]} 

\usepackage{lineno}
\newcommand{\pt}{\mathcal{PT}}
\newcommand{\cspin}{C_{\mathrm{spin}}^2}
\newcommand{\spinlaue}[2]{\ensuremath{[#1\!\parallel\!#2]}}

\usepackage{amsmath, amssymb} 
\usepackage{lineno} 

\usepackage{xcolor}

\newcommand{\bluemark}[1] {\color{black}{#1}\color{black}\normalsize}

\begin{document}

\title{Engineering Altermagnetism via Layer Shifts and Spin Order in Bilayer MnPS$_3$}

\author{J. W. Gonz\'alez}

\affiliation{Departamento de Física, Universidad de Antofagasta, Av. Angamos 601, Casilla 170, Antofagasta, Chile}

\author{T. Brumme}
\affiliation{Chair of Theoretical Chemistry, Technische Universität Dresden, Bergstrasse 66, 01069 Dresden, Germany.}

\author{E. Su\'{a}rez Morell}
\affiliation{Departamento de F\'{i}sica, Universidad 
T\'{e}cnica Federico Santa Mar\'{i}a, Casilla Postal 
110V, Valpara\'{i}so, Chile.}

\author{A. M. León}

\affiliation{Departamento de Física, Facultad de Ciencias, Universidad de Chile, Casilla 653, Santiago, Chile.}
\affiliation{Dresden University of Technology, Institute for Solid State and Materials Physics, 01062 Dresden, Germany\\}


\begin{abstract}
\noindent \hspace{6.3cm}\textbf{Abstract}\\
Altermagnetic materials combine compensated magnetic order with momentum-dependent spin splitting, offering a fundamentally new route for spintronic functionality beyond conventional ferromagnets and antiferromagnets. While most studies have focused on three-dimensional compounds, the emergence of altermagnetism in few-layer two-dimensional materials remains largely unexplored.
Here, we demonstrate that bilayer MnPS$_3$, a prototypical 2D van der Waals magnet, can host stacking-induced altermagnetic phases. Using density-functional theory and spin-Laue symmetry analysis, we show that interlayer spin alignment and lateral displacement act as coupled symmetry control parameters that switch the system between Type II (collinear AFM) and Type III (altermagnetic)  phases.
These results establish stacking engineering as a powerful, purely structural route for designing tunable altermagnetic states in 2D magnets, opening pathways toward symmetry-driven spintronic and magnetoelectronic devices.
\end{abstract}

\maketitle

\section*{Introduction}
\lettrine[lines=2, loversize=0.25, findent=2pt]{\LettrineFont{{A}}}ltermagnetic (AM) materials have emerged as a compelling class of magnetic systems, uniquely combining compensated magnetic order with momentum-dependent spin splitting. This distinctive feature makes them promising candidates for next-generation spintronic applications, offering an alternative to conventional antiferromagnets~\cite{chu2020linear}. While most foundational studies have focused on three-dimensional compounds, the investigation of AM behavior in two-dimensional (2D) systems remains at an early stage~\cite{sun2025proposing,mazin2023induced,liu2024twisted,gonzalez2025altermagnetism}. Importantly, 2D magnets exhibit enhanced sensitivity to external perturbations such as strain, electric fields, and stacking, making them ideal platforms for engineering tunable spin phenomena. Incorporating AM characteristics into 2D materials could enable controllable switching of magnetic states, opening new avenues for functional device design.\\

Extending altermagnetic behavior to two-dimensional few-layer systems has presented several challenges~\cite{mazin2023induced,liu2024twisted,Pan2024}. High structural symmetry tends to protect spin degeneracy: spatial operations such as inversion $P$, out-of-plane mirror $M_z$, or out-of-plane twofold rotation $C_{2z}$ (and their combinations with time reversal) enforce degeneracy and thereby suppress momentum-dependent splitting~\cite{mazin2023induced}. Realizing AM states in van der Waals systems therefore requires breaking those degeneracy-enforcing symmetries while retaining or generating a valid spin-exchanging connector. Bilayers provide a flexible platform because stacking displacement and interlayer spin alignment independently tune which spatial and antiunitary symmetries survive. Recent proposals have invoked layer flipping or twisting to induce AM in antiferromagnetically coupled layers~\cite{Pan2024, liu2024twisted, zeng2024bilayer}; a central requirement in these schemes is the existence of a spatial operation that, when paired with a spin inversion, relates opposite-spin sublattices without itself enforcing degeneracy.
Symmetry analysis identifies two necessary ingredients for Type III altermagnetism: (i) a surviving spin-exchanging connector $\SL{a}{\cspin}$, where the spatial part $a$ is not one of the degeneracy-enforcing operations $\mathcal{P}$, $M_z$, or $C_{2z}$, and (ii) that connector maps opposite-spin sublattices, producing the alternating, momentum-dependent splitting characteristic of altermagnets~\cite{Zeng2024Description, Pan2024, Smejkal2022}. Here, $\cspin$ denotes a $180^\circ$ spin-space rotation that flips the quantization axis (e.g., $S_z \to -S_z$). Pure spatial operations such as $\SL{C_{2z}}{E}$ or $\SL{M_z}{E}$ (or their nonsymmorphic variants) do not qualify as connectors by themselves; they must be accompanied by the spin flip $\cspin$ to lift degeneracy~\cite{Zeng2024Description, Bai2025}.


\begin{figure*}[h!]
\includegraphics[width=0.9\textwidth]{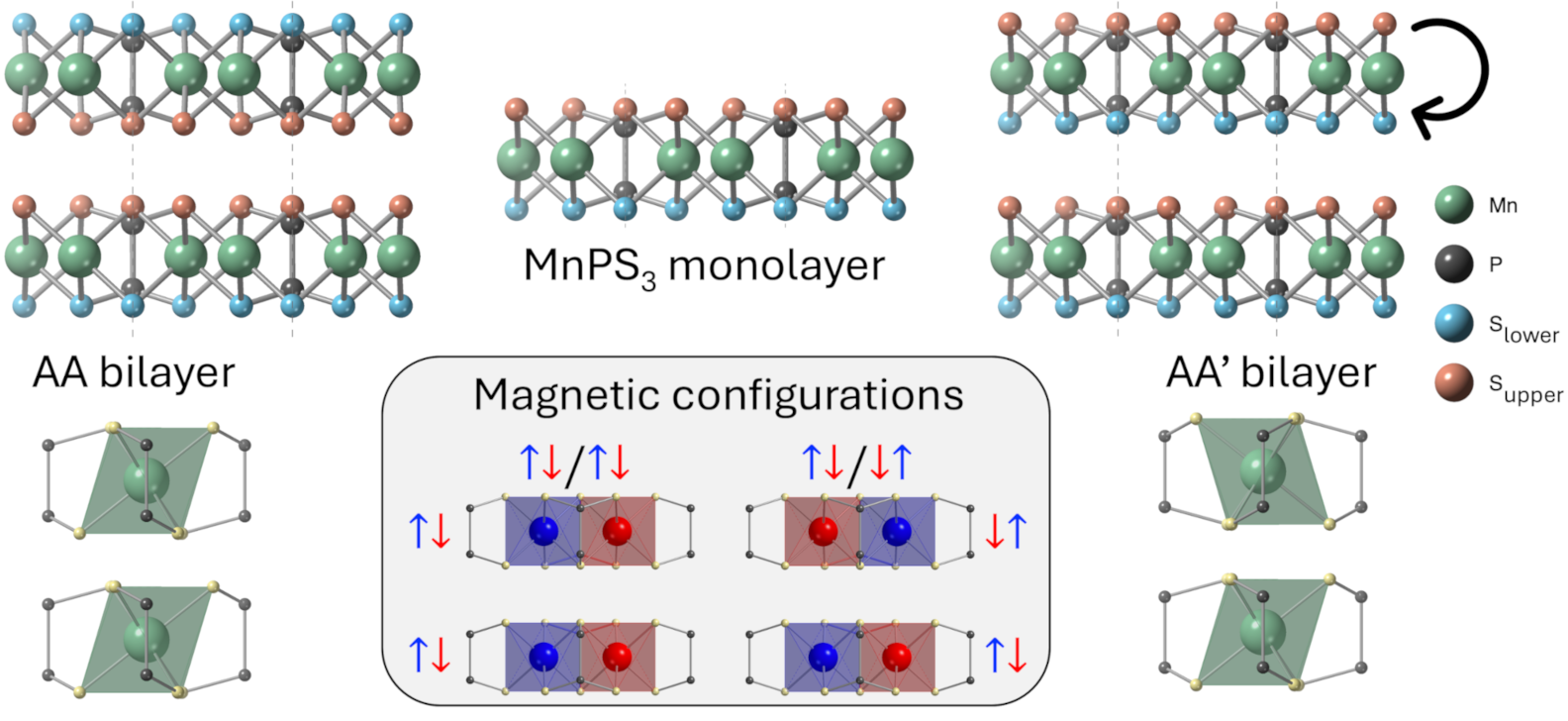}   
\caption{
Atomic and magnetic configurations of MnPS$_3$ monolayers and bilayers.  
\textbf{Top:} Crystal structures of the MnPS$_3$ monolayer, AA and AA$^\prime$ bilayers. 
\bluemark{The two distinct sulfur planes (upper and lower) are highlighted in the monolayer structure to clarify how their relative alignment defines the AA and AA$^\prime$  stacking geometries.}
In the AA$^\prime$ geometry, the upper layer is inverted along the out-of-plane axis ($z \rightarrow -z$). \textbf{Bottom left and right:} Local Mn coordination environments in AA and AA$^\prime$ bilayers, respectively.  
\textbf{Bottom center:} Interlayer magnetic configurations derived from the antiferromagnetic monolayer. The bilayer can adopt either parallel AFM ($\uparrow\downarrow / \uparrow\downarrow$) or antiparallel AFM ($\uparrow\downarrow / \downarrow\uparrow$), each leading to distinct symmetry and magnetic phase behavior.
}
\label{fig:scheme}
\end{figure*}

In this context, TMPX$_3$ MnPX$_3$ (X = S and Se) compounds are attracting increasing attention due to their robust in-plane antiferromagnetic order. Depending on the transition metal (TM), the interlayer spins can align either parallel or antiparallel, giving rise to distinct magnetic couplings and stacking-dependent responses \cite{chu2020linear}. Among them, bilayers \cite{sun2025proposing, Liu2024Robust} stand out as promising candidates for altermagnetism, since their Néel-type structure naturally fulfills the required symmetry constraints \cite{sun2025proposing}. Stacking-dependent studies on related Fe- and Ni-based TMPS$_3$ systems have shown that interlayer shifts can strongly modulate the magnetic coupling \cite{leon2025interlayer}, and similar effects are now being uncovered in MnPX$_3$. In particular, bilayer MnPX$_3$ provides a versatile platform where stacking and parallel     ($\uparrow\downarrow/\uparrow\downarrow$) v/s antiparallel    ($\uparrow\downarrow/\downarrow\uparrow$) interlayer spin alignment (see Fig.~\ref{fig:scheme}) may dictate whether the ground state favors AFM or AM order. Recent work on MnPSe$_3$ further reveals robust AM signatures coupled to sliding ferroelectricity, opening routes for magnetoelectric cross-control \cite{sun2025proposing}. Yet the precise nature of the ground state in MnPX$_3$ bilayers, AFM or AM, remains unsettled across stacking configurations.\\

In this work, we investigate bilayer MnPS$_3$ using first-principles calculations combined with spin-Laue symmetry analysis, a rigorous framework for classifying magnetic phases based on their unitary and antiunitary symmetries~\cite{Liu2024, Zeng2024Description}. By systematically varying stacking and interlayer magnetic order, we reveal the conditions under which Type II (collinear AFM) and Type III (altermagnetic) phases emerge. Our results identify the symmetry elements responsible for preserving or lifting spin degeneracy, establishing stacking and magnetic order as coupled symmetry control parameters for spintronic design in 2D magnets.

\section*{Results}

\begin{figure*}[h!]
\includegraphics[width=0.7\textwidth]{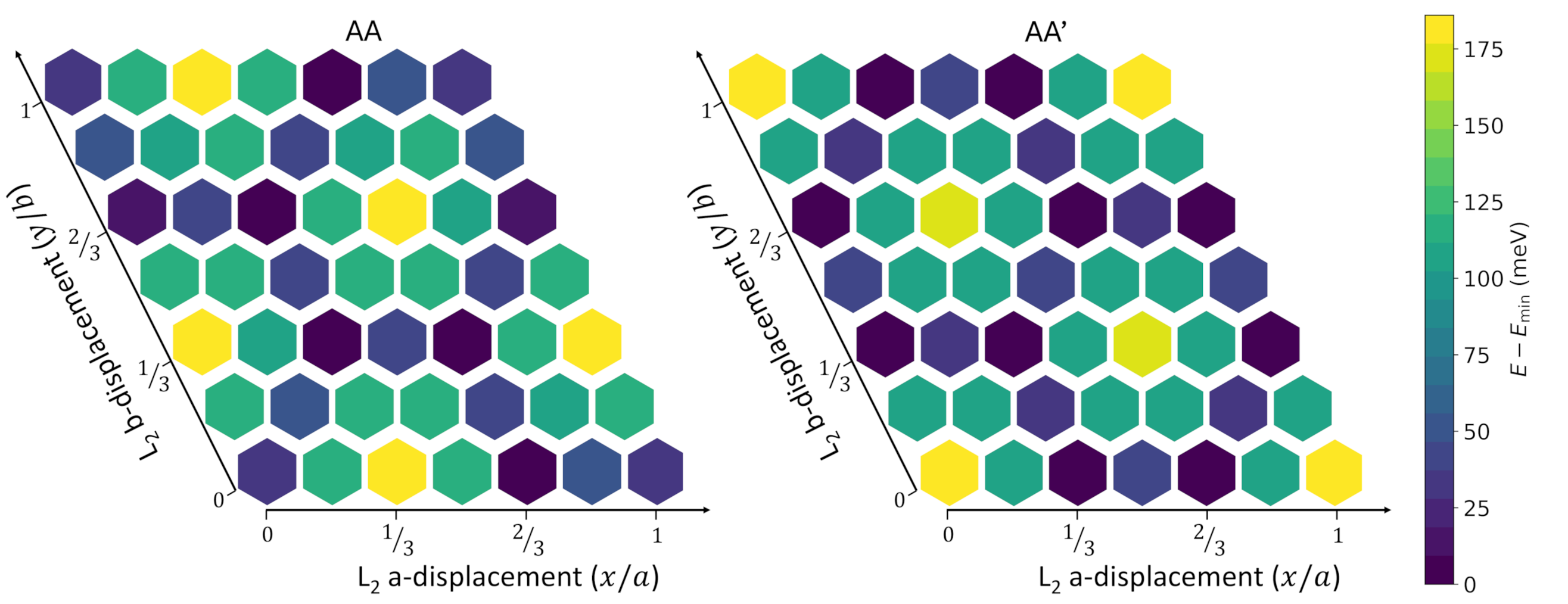}   
\caption{
Relative energy maps for bilayer MnPS$_3$ under lateral displacement in the parallel ($\uparrow\downarrow / \uparrow\downarrow$) magnetic configuration. \textbf{Left:} AA stacking series. \textbf{Right:} AA$^\prime$ stacking series. 
\bluemark{Color scale indicates the relative energy (in meV) for each in-plane displacement, referenced to the lowest-energy configuration found within that stacking series (AA or AA$^\prime$). The global energy minimum of the AA series is 7.67 meV lower than that of the AA$^\prime$ series.}}
\label{fig:ene}
\end{figure*}

\noindent\textbf{Stacking-Dependent Energy Landscape}

We constructed bilayer MnPS$_3$ structures from two fundamental stacking motifs to assess how interlayer configuration influences energetic stability, as schematically illustrated in Fig.~\ref{fig:scheme}. In the AA stacking, the second monolayer is placed directly above the first at the equilibrium interlayer distance $d$, preserving the in-plane atomic registry. In the AA$^\prime$ stacking, the upper layer is first reflected across the out-of-plane axis ($z\to -z$), an operation that, for a centrosymmetric monolayer, is equivalent to a combination of inversion and a 180° in-plane rotation, and then translated vertically by the same distance. Each base structure was subsequently displaced by a lateral vector $\mathbf{t}$, defined as a fractional translation within the $xy$ plane. The resulting energy landscapes appear in Fig.~\ref{fig:ene}.

The AA stacking is energetically preferred: its global minimum lies 7.67 meV per unit cell below the most stable AA$^\prime$ configuration. In the AA landscape, minima occur at high-symmetry lateral displacements such as $(1/3,1/3)$, $(2/3,1/3)$, and $(1/3,2/3)$, corresponding to AB-type registries that preserve an interlayer inversion-like symmetry. In contrast, the AA$^\prime$ profile features several nearly degenerate local minima at commensurate offsets $(n/3,m/3)$, including the so-called AB$^\prime$ configuration at $(1/3,1/3)$~\cite{sun2025proposing}. For example, the configurations at $(0,1/3)$, and $(1/3,0)$ differ in energy by less than 2 meV/cell, illustrating the greater stacking flexibility in AA$^\prime$. Despite that flexibility, AA$^\prime$ remains close in energy to AA, with its average energy only 2.73 meV per unit cell higher.

Importantly, symmetry-equivalent or similar-looking displacements do not guarantee equal energetics. For instance, $(1/3,1/3)$ and $(1/3,2/3)$ nominally correspond to different space groups (e.g., $C2/m$ versus $P\overline{3}$) and differ by 7.21 meV. We attribute such contrasts primarily to the vertical registry of sulfur atoms: configurations in which S atoms sit above metal or P sites tend to minimize interlayer Coulomb repulsion and are energetically favored, whereas those with S atop S enhance repulsion and become higher in energy (highlighted with the yellow hexagon)~\cite{leon2025interlayer}. \bluemark{More details on the energy differences among the high-symmetry stackings are provided in the Supplementary Information. In the following section, we present the analysis for these high-symmetry configurations, while the (1/2,0) stacking is included for comparison.}

\noindent\textbf{Crystallographic Symmetry Analysis} \label{sec:symmetry_analysis}

We analyze how stacking geometry modifies the symmetry of bilayer MnPS$_3$, a central factor in enabling or suppressing nonrelativistic spin splitting. The key distinction is not simply the presence of spatial inversion ($\mathcal{P}$) or time-reversal symmetry ($\mathcal{T}$), but the preservation of their combined antiunitary product $\mathcal{PT}$. This operation can enforce a Kramers-like spin degeneracy across the Brillouin zone in compensated antiferromagnets, even though $(\mathcal{PT})^2 = +1$~\cite{Bai2025, Zeng2024Description,Mavani2025}. We adopt the shorthand $\mathcal{PT} \equiv \SL{\mathcal{P}}{\mathcal{T}}$ throughout the text to refer to this combined symmetry.

While the conventional Kramers theorem requires time-reversal symmetry with $\mathcal{T}^2 = -1$, the antiunitary operator $\mathcal{PT}$ can still protect twofold degeneracies under specific conditions. As shown in related antiferromagnetic systems such as MnPSe$_3$ and MnSe~\cite{Sheoran2024, Qayyum2024}, this protection arises when Bloch states transform as real representations under $\mathcal{PT}$. In collinear antiferromagnets with negligible spin-orbit coupling, $\mathcal{PT}$ maps each Bloch state to an orthogonal partner with opposite spin, enforcing a robust symmetry-protected degeneracy that mimics the Kramers mechanism.
This mechanism is fundamentally tied to the invariance of the system under the combined $\mathcal{PT}$ operation. When this symmetry is preserved, as in centrosymmetric antiferromagnets, spin degeneracy persists at all $\mathbf{k}$-points~\cite{Qayyum2024}. When it is explicitly broken, for instance, due to stacking geometry or interlayer spin orientation, the degeneracy lifts, allowing momentum-dependent spin splitting, a hallmark of altermagnetic behavior~\cite{Sheoran2024,Mavani2025}. The presence or absence of $\mathcal{PT}$ therefore provides a clear symmetry-based distinction between Type II antiferromagnetic phases and the Type III or quasi-altermagnetic regimes.

In AA stacking, both layers share the same in-plane lattice. The survival of $\mathcal{PT}$ depends on the lateral shift $\mathbf{t}$ and the interlayer spin alignment. At high-symmetry offsets such as $(1/3,2/3)$, the nonmagnetic and parallel ($\uparrow\downarrow/\uparrow\downarrow$) magnetic states admit centrosymmetric magnetic space groups (e.g., $P\overline{3}1m'$ (162.75), $P\overline{3}'$ (147.15)) that preserve $\mathcal{PT}$ and protect spin degeneracy. In contrast, antiparallel ($\uparrow\downarrow/\downarrow\uparrow$) alignment always breaks $\mathcal{PT}$. In this case, spin splitting depends on whether a valid spin-exchanging connector survives. The spatial operation $a$ must not belong to the degeneracy-protecting set $\langle \mathcal{P}, M_z, C_{2z} \rangle$, and must combine with spin inversion to form $\SL{a}{C^2_\mathrm{spin}}$. Examples include in-plane twofold rotations such as $C_{2,[110]}$ or mirrors like $m_x$. If such a connector remains, the system exhibits Type III altermagnetism. If no connector survives (e.g., at $\mathbf{t} = (1/3,2/3)$ with tightened symmetry tolerances), the phase becomes quasi-altermagnetic. We provide a detailed symmetry-based definition and diagnostic of the Quasi-AM regime in Sec.~\ref{Laue}.

In AA$^\prime$ stacking, the inversion of the top layer about the $z$ axis ($z \rightarrow -z$) combined with a vertical translation $\tau_z$ reshapes the symmetry landscape, usually eliminating global inversion and thereby breaking $\mathcal{PT}$. However, when $2\mathbf{t} \in \mathcal{L}$, with $\mathcal{L}$ the in-plane Bravais lattice, the composite stacking operation $[\mathcal{P} \parallel \tau_z]$ can combine with time reversal $\mathcal{T}$ to form an effective antiunitary symmetry $[\mathcal{P} \parallel \tau_z]\mathcal{T}$, which enforces spin degeneracy in the parallel alignment~\cite{Pan2024,ramazashvili2009kramers,Zeng2024Description}. In the generic case, however, AA$^\prime$ displacements lower the symmetry to noncentrosymmetric magnetic space groups (MSGs)~\cite{Liu2024,Smolyanyuk2024}, which lack exact $\mathcal{PT}$. As a result, degeneracy protection is lost and Type~III altermagnetism emerges, unless the effective antiunitary symmetry is reconstructed.

\begin{figure*}[h!]
\includegraphics[width=0.7\textwidth]{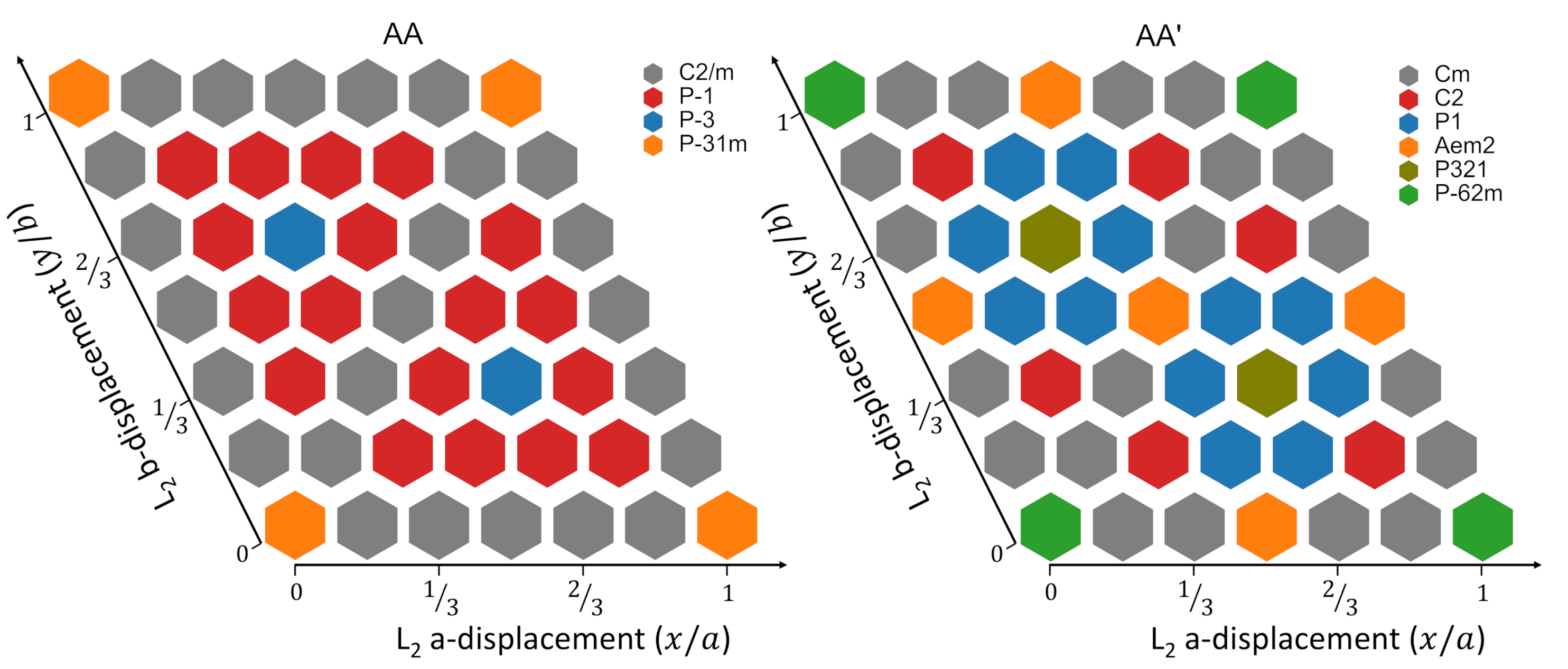}   
\caption{Calculated space group symmetry maps for bilayer MnPS$_3$ under lateral displacement. \textbf{Left:} AA stacking series. \textbf{Right:} AA$^\prime$ stacking series. Color coding indicates the space group symmetry of the bilayer for each relative in-plane displacement vector $\mathbf{t}$ applied to the second layer (L$_2$).}
    \label{fig:sim}
\end{figure*}

Figure~\ref{fig:sim} summarizes the magnetic space groups for different lateral shifts. AA configurations mostly fall under Type-II MSGs in the parallel state, consistent with protected spin degeneracy. In contrast, AA$^\prime$ stackings span lower-symmetry, often noncentrosymmetric groups. Only at special displacements does inversion-like pairing re-emerge to preserve $\mathcal{PT}$.
Although conventional space group assignment offers structural insight, it does not alone determine whether nonrelativistic spin splitting occurs. Altermagnetism depends on an intricate interplay of spatial, spin, and antiunitary symmetries, particularly $\mathcal{PT}$. In many configurations, it is the existence of a valid spin-exchanging connector that dictates whether splitting arises. To expose these subtleties, we adopt the spin-Laue formalism, which explicitly treats symmetries as composite operations in real and spin space, including both $\SL{a}{C^2_\mathrm{spin}}$ and antiunitary elements like $\mathcal{PT}$~\cite{Pan2024, Zeng2024Description, Liu2024}.

\noindent\textbf{Magnetic Configurations of the Bilayer}

We consider two collinear antiferromagnetic alignments in bilayer MnPS$_3$ to isolate how stacking geometry controls magnetic symmetry. Each monolayer hosts the established Néel order on the honeycomb Mn sublattice~\cite{MnPS3_Neel_Monolayer1, MnPS3_Neel_Monolayer2}, with alternating up and down moments. Stacking two such layers produces four Mn sites per unit cell and two distinct interlayer arrangements: in the parallel configuration ($\uparrow\downarrow/\uparrow\downarrow$)  the spin pattern is the same in both layers, while in the antiparallel  ($\uparrow\downarrow/\downarrow\uparrow$) the top layer is flipped relative to the bottom. These magnetic geometries and their relation to stacking are illustrated in Fig.~\ref{fig:scheme}~\cite{leon2025interlayer, Interlayer_Coupling_MPS3_2}. 

Although both alignments are collinear and compensated, they yield distinct spin–Laue structures under lateral displacement, leading to different electronic responses. In all relaxed configurations the net magnetic moment vanishes within numerical tolerance, so the observed spin splitting cannot originate from ferromagnetism but instead reflects symmetry breaking consistent with altermagnetism. The robust intralayer Néel order in MnPS$_3$, stabilized by strong in-plane exchange, is only weakly perturbed by bilayer formation~\cite{leon2025interlayer}. Thus, the two AFM configurations directly link stacking to either symmetry preservation or the emergence of altermagnetic and quasi-altermagnetic spin splitting.

\noindent\textbf{Band Structure}\label{subsec:band_structure}

\bluemark{To explore how stacking-dependent symmetry affects the electronic structure, we analyze the spin-resolved band structures of bilayer MnPS$_3$ for both AA and AA$^\prime$ stackings (Fig.~\ref{fig:bandsAA} and Fig.~\ref{fig:bandsAAp}).

In AA stacking, the parallel spin aliment configuration (Fig.~\ref{fig:bandsAA}, top row) yields fully spin-degenerate bands across all considered in-plane displacements. This spectral regularity directly reflects the preservation of key symmetry elements under such shifts. As shown in Fig.~\ref{fig:sim} (left panel), displacements like $(1/3,2/3)$ and $(1/2,0)$ maintain inversion or vertical mirror symmetries, which constrain the band structure and enforce  identical dispersion for both spin channels throughout the Brillouin zone.
In contrast, the antiparallel configuration (Fig.\ref{fig:bandsAA}, bottom row) exhibits a more varied $\mathbf{k}$-dependent spin-splitting response. The lateral displacement $\mathbf{t}$ directly determines which spatial symmetries are preserved or broken (see Fig.\ref{fig:sim}, left panel). For instance, in the AA+$(1/3,2/3)$ stacking, the antiparallel spin alignment removes global inversion symmetry $\mathcal{P}$, yielding a noncentrosymmetric magnetic space group ($P3$) with no surviving spin-exchanging connectors. This symmetry-starved regime produces significant spin splitting near $\Gamma$ and K$'$, consistent with the quasi-altermagnetic behavior described in recent symmetry-based classifications~\cite{Zeng2024Description,Pan2024}. Overall, AA-based bilayers demonstrate that systematic shifts in stacking registry, combined with interlayer spin order, act as a powerful symmetry control knob for generating tunable spin-resolved features in the electronic structure.

\begin{figure*}[h!]
\includegraphics[width=0.95\textwidth]{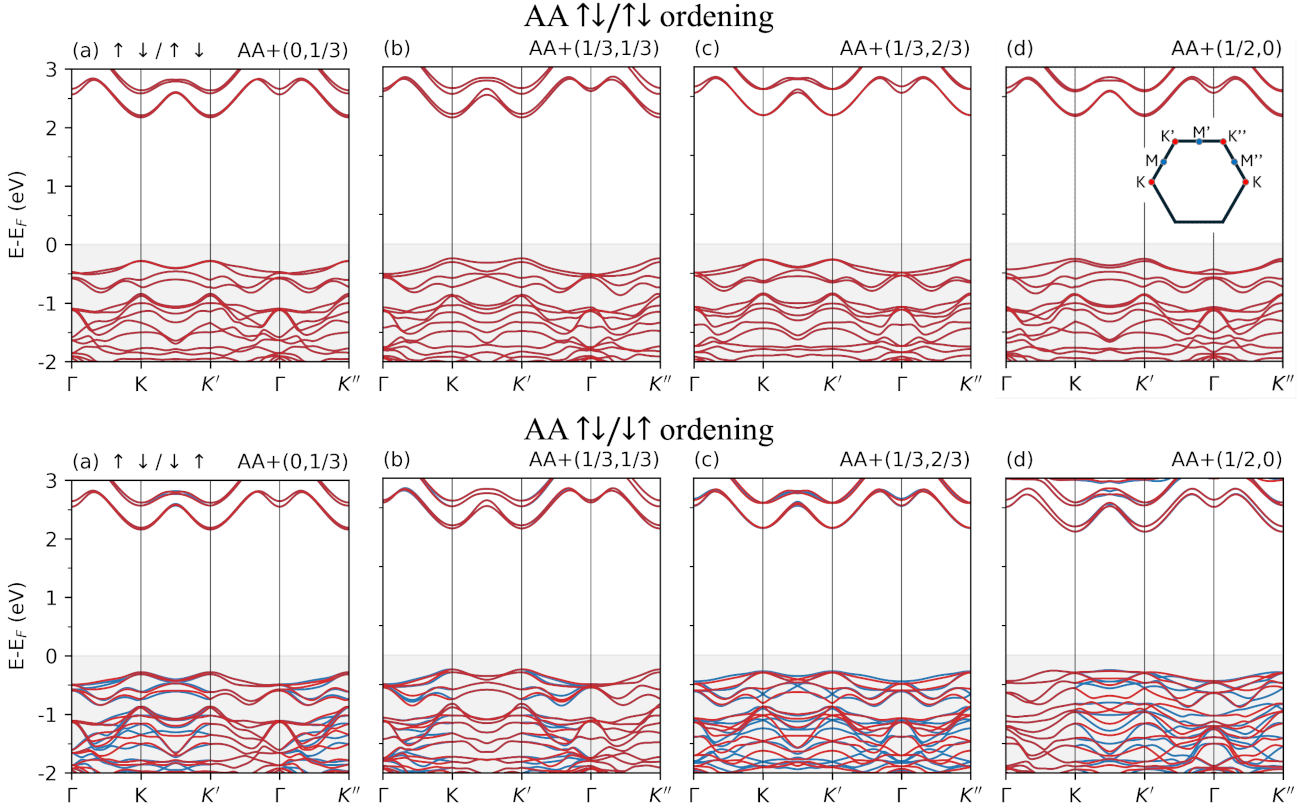}  
\caption{Spin-resolved electronic band structures of AA-stacked bilayer MnPS$_3$ for four in-plane displacements of the top layer: (a) $(0,1/3)$, (b) $(1/3,1/3)$, (c) $(1/3,2/3)$, and (d) $(1/2,0)$, given in fractional lattice coordinates. \textbf{Top row} shows the parallel ($\uparrow\downarrow/\uparrow\downarrow$) and \textbf{bottom row} the antiparallel ($\uparrow\downarrow/\downarrow\uparrow$) interlayer spin configuration. 
Red and blue indicate spin-up and spin-down bands, with the Fermi energy set at $E_F = 0$.}
\label{fig:bandsAA}
\end{figure*}

The AA$^\prime$ stacking geometry intrinsically breaks out-of-plane symmetry due to the reflection of the top layer along $z \rightarrow -z$. As a result, the system becomes more sensitive to in-plane displacements, with Fig.~\ref{fig:sim} (right panel) showing that most lateral shifts reduce the symmetry to lower space groups lacking inversion and mirrors.
This reduction in symmetry is directly manifested in the spin-resolved band structures of Fig.~\ref{fig:bandsAAp}. In the parallel configuration (top row), configurations like AA$^\prime$+$(1/3,2/3)$ and $(1/2,0)$, which retain partial symmetries, show relatively symmetric bands and weak spin splitting. In contrast, for displacements such as $(1/3,1/3)$ and $(0,1/3)$, where Fig.~\ref{fig:sim} indicates the absence of inversion or mirror symmetry, the bands become visibly asymmetric between spin channels, with noticeable differences in curvature and extrema.
The antiparalell configuration (Fig.~\ref{fig:bandsAAp}, bottom row) exhibits robust spin splitting across all displacements, being consistent with the lack of any global or partial symmetry connecting the spin sublattices, as confirmed by the stacking-dependent space group analysis.  These results establish a clear correspondence between lateral displacement, symmetry reduction, and the emergence of momentum-dependent spin polarization in the band structure.

Notably, the band path includes three non-equivalent K points (K, K$'$, and K$''$), allowing us to resolve anisotropic spin splitting in $k$-space. 
In altermagnetic configurations, spin splitting is selectively observed along only two of the three $\Gamma$-K segments, reflecting the momentum-filtered action of residual symmetry operations.
Section \ref{sec:bnd-split} provides a momentum-resolved analysis of the spin splitting, computed by evaluating the energy difference between spin channels across the Brillouin zone. The combined evolution of spin degeneracy, gap character, and band extrema demonstrates the tunability of stacking and magnetism as a platform for designing spin-functional van der Waals materials.}

The stability of each magnetic configuration and stacking is discussed in Supplementary Information.

\begin{figure*}[ht]
\includegraphics[width=0.95\textwidth]{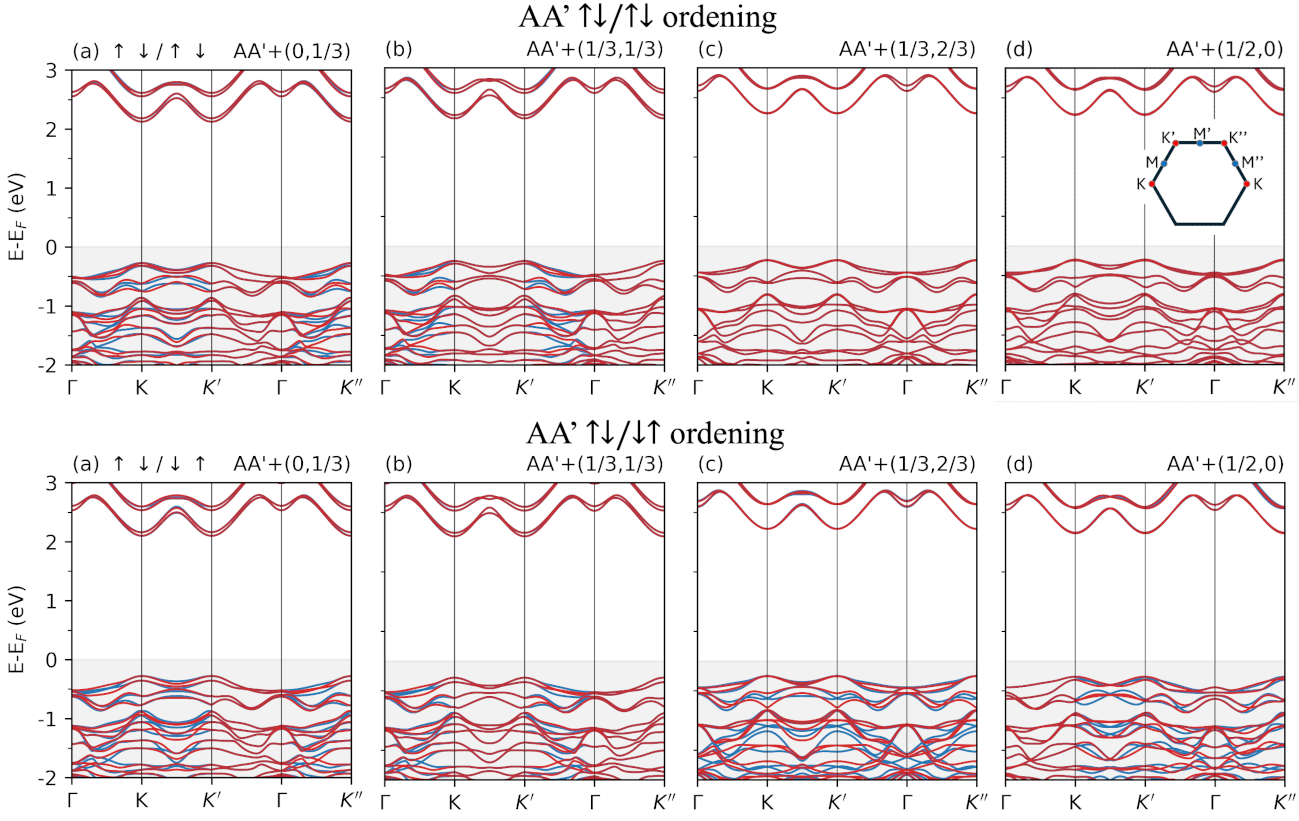}  
\caption{Spin-resolved electronic band structures of AA$^\prime$-stacked bilayer MnPS$_3$ for different in-plane displacements of the top layer. Panel layout mirrors Fig.~\ref{fig:bandsAA}: \textbf{Top row} shows the parallel ($\uparrow\downarrow/\uparrow\downarrow$) and \textbf{bottom row} the antiparallel ($\uparrow\downarrow/\downarrow\uparrow$) interlayer spin configuration. Red and blue denote spin-up and spin-down bands, with the Fermi energy set to $E_F = 0$.
}
\label{fig:bandsAAp}
\end{figure*}

\noindent\textbf{Stacking-Dependent Band Splitting \label{sec:bnd-split}}

\begin{figure*}[h!]
\includegraphics[width=0.85\textwidth]{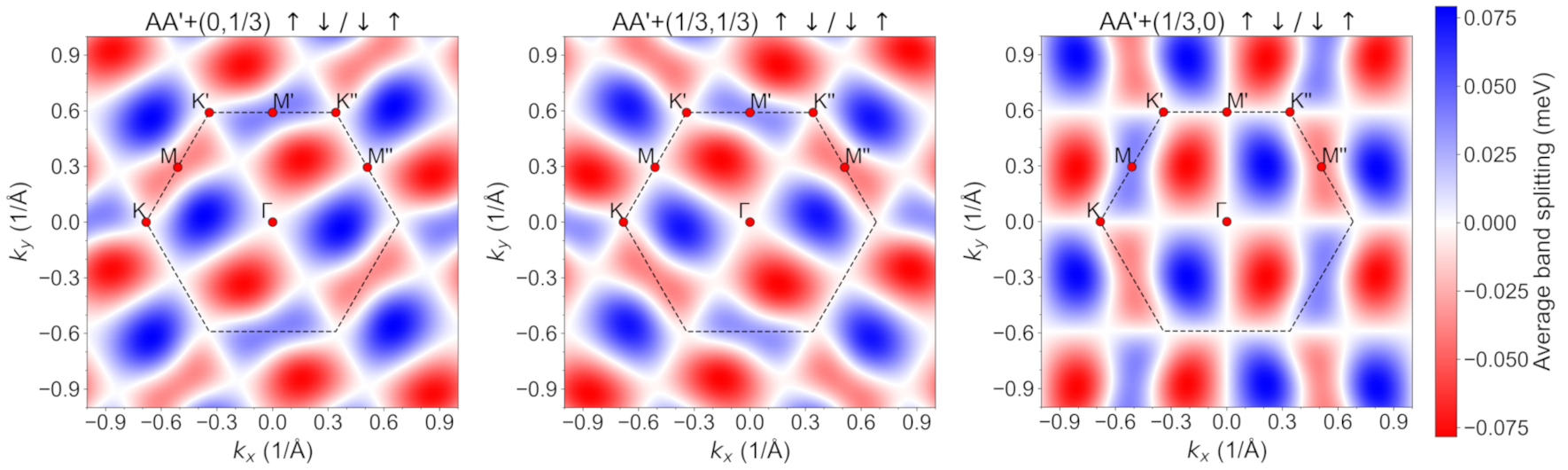}   
\caption{Average spin splitting $\delta E(\mathbf{k})$ (meV) over occupied bands, for AA$^\prime$-stacked MnPS$_3$ bilayers in the antiparallel ($\uparrow\downarrow/\downarrow\uparrow$) ordering with magnetic space group \textit{Cm}, shown in the 2D Brillouin zone. From left to right: lateral displacements $(0,1/3)$, $(1/3,1/3)$, and $(1/3,0)$. Dashed lines mark the hexagonal zone boundary; high-symmetry points are labeled. Positive (blue) values correspond to $E_{\uparrow}>E_{\downarrow}$, negative (red) to $E_{\downarrow}>E_{\uparrow}$, so the sign encodes the direction of spin polarization. The orientation and selection of nonzero splitting reflect the mirror symmetries preserved in each stacking configuration.}
\label{fig:split}
\end{figure*}

To quantify spin splitting in momentum space, we compute the average energy difference between spin-up and spin-down states across the $N$ occupied bands at each $\mathbf{k}$-point:
\begin{equation}
\delta E(\mathbf{k}) = \frac{1}{N} \sum_{n=1}^{N} \left[ E_{\mathrm{up}}(\mathbf{k}, n) - E_{\mathrm{dn}}(\mathbf{k}, n) \right],
\end{equation}
where $E_{\mathrm{up}}$ and $E_{\mathrm{dn}}$ are spin-resolved eigenvalues from self-consistent DFT calculations. This function, $\delta E(\mathbf{k})$, serves as a local measure of spin polarization in reciprocal space \cite{leon2025strain}. Interpolated over a dense $k$-grid, it yields the spin-splitting maps shown in Fig.~\ref{fig:split}, which provide complementary insight to the spin-resolved band structures in Fig.~\ref{fig:bandsAAp}.

Figure~\ref{fig:split} focuses on three low-energy AA$^\prime$ configurations with antiparallel spin  alignment, lateral displacements $(0,1/3)$, $(1/3,1/3)$, and $(1/3,0)$, that all share the monoclinic space group Cm (Fig.~\ref{fig:sim}) but differ in their active mirror content. These subtle differences control the orientation of the spin-splitting texture. For $(0,1/3)$, the mirror
\begin{equation}
M_{(0,1/3)} = \begin{pmatrix} -1 & 0 \\ -1 & 1 \end{pmatrix}
\end{equation}
enforces reflection symmetry about $k_y$; for $(1/3,0)$,
\begin{equation}
M_{(1/3,0)} = \begin{pmatrix} 1 & -1 \\ 0 & -1 \end{pmatrix}
\end{equation}
imposes symmetry about $k_x$; and for $(1/3,1/3)$,
\begin{equation}
M_{(1/3,1/3)} = \begin{pmatrix} 0 & 1 \\ 1 & 0 \end{pmatrix}
\end{equation}
exchanges $k_x$ and $k_y$, producing a diagonal reflection. All three configurations exhibit alternating (checkerboard-like) spin polarization, but the patterns rotate or reflect by their respective mirror constraints. The three textures are related approximately by combinations of the underlying hexagonal lattice rotations and reflections, illustrating that lateral shifts tune not only the magnitude but also the orientation of the anisotropic spin splitting.

Because AA$^\prime$ stacking generally breaks exact inversion-related pairing, it relaxes valley equivalences (e.g., between K, K$'$, and K$''$) that would otherwise be enforced, allowing distinct splitting profiles in different valleys. The surviving or broken connector symmetries further select which $\Gamma\!-\!K$ directions carry nonzero splitting: in true altermagnetic regimes, the splitting appears only along a subset of these segments, reflecting the momentum-filtered action of the residual $\SL{a}{\cspin}$ symmetries.

To illustrate the key features of the altermagnetic phase, Fig.~\ref{fig:split} showcases the spin-splitting maps for three degenerate, low-energy AA$^\prime$ configurations with $\uparrow\downarrow/\downarrow\uparrow$ alignment. These structures, with lateral displacements of (0,1/3), (1/3,1/3), and (1/3,0), all share the monoclinic space group Cm (Fig.~\ref{fig:sim}) but differ in their in-plane mirror content. These subtle symmetry differences dictate the orientation of the spin-splitting textures, showing that a simple lateral displacement controls both the magnitude and the orientation of the spin anisotropy in momentum space. This stacking-controlled functionality provides the foundation for engineering spatially modulated altermagnetic textures in moiré heterostructures. In twisted bilayers of antiferromagnetic van der Waals materials, the moiré superlattice generates a periodic array of local stacking registries, each defined by a distinct symmetry environment. As demonstrated by our results, such a moiré pattern can host coexisting altermagnetic and quasi-altermagnetic domains, where each domain exhibits a distinct momentum-dependent spin texture governed by its local stacking displacement $\mathbf{t}$. This perspective is consistent with recent theoretical proposals for twist-controlled spin splitting in MnPSe$_3$~\cite{Sheoran2024} and with the broader potential of moiré engineering in magnetic van der Waals materials~\cite{Pan2024}. Interfaces between these domains may support unconventional phenomena, such as spin-polarized edge states or emergent gauge fields arising from spatially varying symmetry constraints. Exploring such domain boundaries can therefore bridge the physics of altermagnetism with moiré engineering, paving the way for spatially programmable, symmetry-defined spin functionalities in compensated magnets.

\noindent\textbf{Layer and Laue Spin Groups\label{Laue}}

The classification of magnetic phases in two-dimensional van der Waals magnets requires a framework that distinguishes between symmetry-protected spin degeneracy, momentum-dependent spin splitting, and unprotected splitting arising from broken symmetry. We adopt the spin-Laue formalism~\cite{Liu2024, Zeng2024Description}, expressing magnetic symmetry operations as combined spatial and spin actions $\SL{R_{\mathrm{spatial}}}{S_{\mathrm{spin}}}$, where $R_{\mathrm{spatial}}$ acts on lattice coordinates and $S_{\mathrm{spin}}$ on spin. This classification yields an unambiguous taxonomy of compensated collinear magnets according to which symmetries survive:

\textbf{Type II Antiferromagnet:} A degeneracy-enforcing antiunitary symmetry, most prominently the combined inversion-time reversal operation $\mathcal{PT}$, remains. When this symmetry commutes with the Hamiltonian and the system supports real or pseudoreal electronic states, it guarantees spin degeneracy at all $\mathbf{k}$-points, even though $(\mathcal{PT})^2 = +1$, in contrast to the conventional $\mathcal{T}^2 = -1$ case~\cite{Bai2025, Watanabe2020PRX}. This degeneracy, while not guaranteed by the standard Kramers theorem, emerges as a symmetry-enforced consequence of the compensated magnetic structure and centrosymmetric arrangement. The magnetic space group must therefore be centrosymmetric or contain an equivalent composite operation that enforces such a degeneracy.\\
\textbf{Type III Altermagnet:} $\pt$ is broken, but a spin-exchanging connector $\SL{a}{\cspin}$ survives, where the spatial part $a$ is \emph{not} a degeneracy-enforcing operation such as inversion $\mathcal{P}$, out-of-plane mirror $M_z$, or out-of-plane twofold rotation $C_{2z}$. Examples include in-plane twofold rotations (e.g., $C_{2,[110]}$) or in-plane mirrors (e.g., $m_x$). Such $a$, when paired with the spin inversion $\cspin$, relates opposite-spin sublattices and imposes the alternating, momentum-dependent spin splitting characteristic of altermagnetism~\cite{Zeng2024Description, Pan2024}.\\
\textbf{Quasi-Altermagnet (Quasi-AM):} 
This phase constitutes a specific subclass within the SST-4 class of compensated antiferromagnets, characterized by nonrelativistic spin splitting at time-reversal-invariant momenta such as $\Gamma$, in the absence of spin-orbit coupling~\cite{yuan2024nonrelativistic}. 
Unlike Type III altermagnets, the Quasi-AM phase lacks both the combined antiunitary symmetry $\mathcal{PT}$ and any valid spin-exchanging connector $\SL{a}{\cspin}$, where the spatial operation $a$ lies outside the degeneracy-enforcing set $\langle \mathcal{P}, M_z, C_{2z} \rangle$. The absence of such symmetries implies that the two opposite-spin sublattices are not related by any preserved operation of the magnetic space group, leading to a complete lifting of spin degeneracy even at $\Gamma$. The resulting spin splitting is unprotected and does not exhibit the symmetry-enforced alternating sign in momentum space that characterizes true altermagnets. 

Before presenting the symmetry-based classification of stacking geometries, we clarify the criteria that determine whether a spin-exchanging connector $\SL{a}{\cspin}$ is operative. Such a connector is valid only if the spatial operation $a$ does not generate a degeneracy-enforcing antiunitary symmetry when combined with time reversal $\mathcal{T}$. In particular, operations like $\mathcal{P}$, $M_z$, and $C_{2z}$ belong to the set $\langle \mathcal{P}, M_z, C_{2z} \rangle$, which can protect spin degeneracy when preserved alongside $\mathcal{T}$~\cite{Zeng2024Description,Smejkal2022,ramazashvili2009kramers}. 
The symmetry analysis is framed in terms of the magnetic point group $H$, defined as the set of spatial operations that preserve both the atomic structure and the magnetic configuration. Each element $h \in H$ corresponds to a symmetry operation (such as a rotation, mirror, or glide) that leaves the spin texture invariant. These generate spin-space operations of the form $[h \parallel S]$, where $S$ acts on the spin degrees of freedom. The subgroup $\{[h \parallel E] : h \in H\}$, with $E$ the identity in spin space, preserves local spin orientation and forms the spin-conserving core of the spin-Laue group~\cite{Liu2024,Smolyanyuk2024}.

In contrast, a spin-exchanging connector $\SL{a}{\cspin}$ flips the spin while the spatial part $a$ maps spin-up sites to spin-down counterparts. For example, although $C_{2z} \in \langle \mathcal{P}, M_z, C_{2z} \rangle$, the connector $[C_{2z} \parallel \cspin]$ becomes operative only if the composite antiunitary symmetry $[C_{2z} \parallel \mathcal{T}]$ is broken. This occurs in antiparallel interlayer alignments ($\uparrow\downarrow/\downarrow\uparrow$), where $C_{2z}$ maps spin-up sites to spin-down sites in a manner incompatible with time reversal. When this condition is met, spin degeneracy is lifted, and alternating momentum-dependent spin splitting emerges, characteristic of Type III altermagnetism\cite{Smejkal2022}.

The specific composition of the spin-Laue group determines the resulting magnetic phase. A purely spin-conserving group produces spin-degenerate bands (Type II AFM); the inclusion of valid spin-exchanging connectors leads to alternating spin splitting (Type III); and the absence of either symmetry channel results in unprotected splitting (Quasi-AM)\cite{yuan2024nonrelativistic}. All magnetic space groups reported below are derived from fully relaxed structures (forces $<10^{-3}$ eV/\AA), using strict symmetry tolerances. Phase labels are cross-validated against DFT band structures, which directly reveal whether spin degeneracy is preserved or lifted.\\

\noindent\textbf{Spin–Laue Classification: AA Stacking}

AA stacking provides a tunable framework for controlling spin degeneracy in bilayer MnPS$_3$, since lateral displacement $\mathbf{t}$ and interlayer spin alignment act as largely independent symmetry control parameters~\cite{Pan2024,fox2023stacking}. In AA-stacked MnPS$_3$, the magnetic phase is dictated by the presence of $\mathcal{PT}$ or, when absent, by the survival of a valid connector $\SL{a}{\cspin}$ across spin sublattices~\cite{Zeng2024Description,Smejkal2022}. 
We identify the magnetic space group  (MSG) of each configuration to determine which symmetries remain after stacking and magnetic ordering~\cite{Liu2024}. The resulting classification appears in Table~\ref{tab:AA_laue}.

In the parallel alignment ($\uparrow\downarrow/\uparrow\downarrow$) the combined inversion–time reversal $\mathcal{PT}$ persists for all tested offsets~\cite{ramazashvili2009kramers,zeng2024bilayer}. Although the MSG varies with $\mathbf{t}$ (e.g., $P\overline{3}1m'$ at $(0,0)$, $C2'/m'$ at $(0,1/3)$ or $(1/2,0)$, and $P\overline{3}'$ at $(1/3,2/3)$), the centrosymmetric character guarantees the survival of $\mathcal{PT}$ enforces a Kramers-like spin degeneracy throughout the Brillouin zone. Hence all parallel AA configurations are Type II AFMs, as confirmed by the absence of intrinsic splitting in DFT~\cite{mazin2023induced,liu2024twisted}.

The antiparallel alignment ($\uparrow\downarrow/\downarrow\uparrow$) breaks $\mathcal{PT}$ for any $\mathbf{t}$, so splitting depends on connector survival. At high-symmetry offsets such as $(0,0)$, spatial operations like $C_{2z}$ and $C_{2,[110]}$, when paired with $\cspin$, yield valid connectors $\SL{C_{2z}}{\cspin}$ and $\SL{C_{2,[110]}}{\cspin}$, producing the alternating, momentum-dependent splitting of a Type III altermagnet~\cite{Zeng2024Description,Altermagnetism_Pattern1,rooj2025altermagnetic}. Similarly, at $(0,1/3)$ or $(1/2,0)$ surviving mirrors (e.g., $m_x$) combine with $\cspin$ to give $\SL{m_x}{\cspin}$ connectors and altermagnetic behavior, with the expected vanishing at symmetry-imposed nodes~\cite{Bai2025,Sheoran2024}.

The offset $\mathbf{t} = (1/3, 2/3)$ presents a distinct case. Under antiparallel alignment, detailed symmetry analysis with tightened numerical thresholds confirms the absence of a valid inversion pairing required for preserving $\mathcal{PT}$~\cite{Smolyanyuk2024}. No spin-exchanging connector $\SL{a}{\cspin}$ survives either. The resulting magnetic space group reduces to noncentrosymmetric $P3$ (No.~144), leading to unprotected spin splitting, even at $\Gamma$ (see, Fig.\ref{fig:bandsAA}(c) lower panel). This symmetry-starved configuration fits the SST-4 class and is best described as a Quasi-Altermagnet ~\cite{yuan2024nonrelativistic}.

The progression from a protected Type II AFM through structured Type III altermagnets to a Quasi-AM under AA stacking demonstrates predictable phase selection via combined control of interlayer registry and spin alignment.\\

\begin{table*}[h!]
\centering
\caption{Spin–Laue classification for AA-stacked bilayer MnPS$_3$ under parallel ($\uparrow\downarrow/\uparrow\downarrow$) and antiparallel ($\uparrow\downarrow/\downarrow\uparrow$) configurations.   Parenthetical 2D-AM labels are from Zeng \emph{et al.}~\cite{Zeng2024Description}. The magnetic space groups (MSG) follow the classification in Ref.~\cite{Liu2024}. Connectors are explicit and oriented (e.g., $C_{2z}$ vs in-plane $C_{2,[110]}$ or mirror $m_x$).}
\label{tab:AA_laue}
\vspace{0.5em}

\textbf{AA $\uparrow\downarrow/\uparrow\downarrow$ ordering} \\
\begin{tabular}{@{} c | l | l | l | p{0.3\textwidth} @{}}
\toprule
Offset & Phase (Zeng et al.) & MSG & Spin–Laue Group & Generators / Comment \\
\midrule
(0,0) & AFM (Type II) & $P\overline{3}1m'$ (162.75) &
$\{[h\!\parallel\!E]:h\in H\}\;\cup\;\SL{P}{T}$ &
$H=\overline{3}m$; $\pt$ intact; centrosymmetric; no splitting. \\
\hline
(0,1/3) & AFM (Type II) & $C2'/m'$ (12.60) &
Same as above &
$H=2/m$; $\pt$ preserved. \\
\hline
(1/3,1/3) & AFM (Type II) & $C2'/m'$ (12.60) &
Same as above &
Same as (0,1/3). \\
\hline
(1/3,2/3) & AFM (Type II) & $P\overline{3}'$ (147.15) &
Same as above &
$\pt$ exact; inversion preserved; no splitting. \\
\hline
(1/2,0) & AFM (Type II) & $C2'/m'$ (12.60) &
Same as above &
Same as (0,1/3). \\
\bottomrule
\end{tabular}

\vspace{1em}
\textbf{AA $\uparrow\downarrow/\downarrow\uparrow$  ordering}\\  
\begin{tabular}{@{} c | l | l | l | p{0.3\textwidth} @{}}
\toprule
Offset & Phase (Zeng et al.) & MSG & Spin–Laue Group & Generators / Comment \\
\midrule
(0,0) & AM (Type III / ${}^{1}\overline{3}^{2}m$) & $P31m$ (162.73) & 
$\{[h\!\parallel\!E]:h\in H\}$ & 
$\pt$ broken; both $[C_{2z}\!\parallel\!\cspin]$  \\
  &   &  & 
$\cup\;\{ \SL{C_{2z}}{\cspin}, \SL{C_{2,[110]}}{\cspin} \}$ &  $[C_{2,[110]}\!\parallel\!\cspin]$ act as connectors, enforcing alternating splitting. \\
\hline
(0,1/3) & AM (Type III / ${}^{2}m^{2}m^{1}m$) & $C2/m'$ (12.58) &
$\{[h\!\parallel\!E]:h\in H\}\;\cup\;\{ \SL{m_x}{\cspin} \}$ &
$\pt$ broken; mirror connector $[m_x\!\parallel\!\cspin]$ present. \\
\hline
(1/3,1/3) & AM (Type III / ${}^{2}m^{2}m^{1}m$) & $C2/m'$ (12.58) &
Same as above & Same as (0,1/3). \\
\hline
(1/3,2/3) & Quasi-AM (SST-4) & $P3$ (144) &
$\{[h\!\parallel\!E]:h\in H\}$ &
$\pt$ broken; no valid $\SL{a}{\cspin}$ survives; unprotected splitting at $\Gamma$. \\
\hline
(1/2,0) & AM (Type III / ${}^{2}m^{2}m^{1}m$) & $C2/m'$ (12.58) &
Same as (0,1/3) &
$\pt$ broken; mirror-type connector survives. \\
\bottomrule
\end{tabular}

\begin{flushleft}
\textit{Note:} Parenthetical 2D-AM labels are from Zeng \emph{et al.}~\cite{Zeng2024Description}. 
Type II AFM phases retain degeneracy via $\mathcal{PT}$; 
Type III altermagnets require surviving non-degeneracy-enforcing connectors $\SL{a}{\cspin}$ with $a \notin \{P, M_z, C_{2z}\}$~\cite{Smejkal2022,Zeng2024Description}; 
Quasi-AM phases lack both \cite{yuan2024nonrelativistic}. 
\end{flushleft}
\end{table*}

\noindent\textbf{Spin–Laue Classification: AA$^\prime$ Stacking}

AA$^\prime$ stacking modifies the symmetry environment of bilayer MnPS$_3$ by inverting the top monolayer about its center and translating it vertically by $\tau_z$. This defines the composite spatial operation $[P \parallel \tau_z]$, absent in the monolayer but allowed in the bilayer for specific in-plane displacements $\mathbf{t}$. When $2\mathbf{t} \in \mathcal{L}$, with $\mathcal{L}$ the in-plane Bravais lattice, the antiunitary symmetry $[P \parallel \tau_z]T$ becomes exact and enforces a Kramers-like degeneracy by mapping each Bloch state to an orthogonal partner with opposite spin, even though $([P \parallel \tau_z]T)^2 = +1$~\cite{ramazashvili2009kramers,Pan2024,Zeng2024Description}. In the parallel interlayer configuration ($\uparrow\downarrow/\uparrow\downarrow$), this degeneracy is preserved, stabilizing a Type II antiferromagnetic phase with spin-degenerate bands~\cite{Smejkal2022,zeng2024bilayer}. In contrast, the antiparallel configuration ($\uparrow\downarrow/\downarrow\uparrow$) breaks $[P \parallel \tau_z]T$, lifting the degeneracy and enabling momentum-dependent spin splitting characteristic of quasi-altermagnetic phases~\cite{Sheoran2024,mazin2023induced}. 
We identify the magnetic space group of each AA$^\prime$ configuration to determine the symmetries that survive after stacking and magnetic ordering. The full classification appears in Table~\ref{tab:AAp_laue}.


Magnetic phase outcomes depend on both interlayer spin alignment and lateral shift $\mathbf{t}$. In the antiparallel alignment ($\uparrow\downarrow/\downarrow\uparrow$), the exact $\mathcal{PT}$ is broken for all $\mathbf{t}$, since layer inversion maps spin-up sites to spin-down sites in a way that time reversal alone cannot compensate. Nonetheless, specific shifts retain non-degeneracy-enforcing spatial operations that, when paired with the spin inversion $\cspin$, act as valid connectors. For instance, at $(0,0)$ and $(1/3,2/3)$ the twofold rotation $C_{2z}$ survives and yields $\SL{C_{2z}}{\cspin}$, while at $(0,1/3)$ and $(1/3,1/3)$ in-plane mirrors such as $m_x$ produce $\SL{m_x}{\cspin}$. These connectors enforce the alternating, momentum-dependent splitting of Type III altermagnetism, so the antiparallel AA$^\prime$ stack generally hosts robust Type III AM phases, as corroborated by DFT.

The parallel alignment ($\uparrow\downarrow/\uparrow\downarrow$) presents a different scenario. Layer inversion does not automatically forbid a degeneracy-enforcing composite symmetry, but restoring an effective $\mathcal{PT}$ requires geometric compatibility: only when $2\mathbf{t}\in\mathcal{L}$ (with $\mathcal{L}$ the in-plane Bravais lattice) does the combination of layer inversion and translation reconstruct an inversion pairing that properly relates opposite-spin sublattices. At the special offset $\mathbf{t}=(1/2,0)$ this condition holds, the system exhibits spin-degenerate bands, and the phase behaves as a Type II AFM; the spatial group is $Aem2$ (39), which supports the effective 
$\mathcal{PT}$. For shifts with $2\mathbf{t}\notin\mathcal{L}$, the effective inversion fails and if a valid spin-exchanging connector survives (e.g., $\SL{C_{2z}}{\cspin}$ in appropriate cases), the system enters the Type III altermagnetic regime.

This contrasts with AA stacking, where the parallel case universally preserves $ \mathcal{P} \mathcal{T}$ and the antiparallel case requires connector survival to realize altermagnetism or, in its absence, a Quasi-AM. In AA$^\prime$, the internal inversion renders Type III altermagnetism the default for antiparallel order, while the parallel alignment permits a controlled reentry into a Type II AFM only at specific high-symmetry registries. Thus stacking geometry provides a distinct symmetry control pathway: intrinsic symmetry breaking stabilizes altermagnetism except when precise registry restores an effective degeneracy-enforcing composite symmetry.

\begin{table*}[ht!]
\centering
\caption{Spin–Laue classification for AA$^\prime$-stacked bilayer MnPS$_3$ under parallel and antiparallel magnetic orderings. Parenthetical 2D-AM labels are from Zeng \emph{et al.}~\cite{Zeng2024Description}. The magnetic space groups (MSG) follow the classification in Ref.~\cite{Liu2024}.} 
\label{tab:AAp_laue}
\vspace{0.5em}

\textbf{AA$^\prime$ $\uparrow\downarrow/\uparrow\downarrow$ (Parallel) ordering} \\
\begin{tabular}{@{} c | l | l | l | p{0.30\textwidth} @{}}
\toprule
Offset / Structure & Phase (Zeng et al.) & MSG & Spin–Laue Group & Generators / Comment \\
\midrule
(0,0) & AM (Type III / ${}^{1}\overline{3}^{2}m$) & $P\overline{6}'2'm$ (189.221) & 
$\{[h\!\parallel\!E]:h\in H\}\;\cup\;\spinlaue{C_2}{\cspin}$ &
Broken $\pt$; $[C_2\!\parallel\!\cspin]$ connector present. \\
\hline
(0,1/3) & AM (Type III / ${}^{2}2/^{2}m_x$) & $C'm$ (8.32) & 
$\{[h\!\parallel\!E]:h\in H\}\;\cup\;\spinlaue{m_x}{\cspin}$ &
Broken $\pt$; mirror-type connector $[m_x\!\parallel\!\cspin]$. \\
\hline
(1/3,1/3) & AM (Type III / ${}^{2}2/^{2}m_x$) & $C'm$ (8.32) &
$\{[h\!\parallel\!E]:h\in H\}\;\cup\;\spinlaue{m_x}{\cspin}$ &
Broken $\pt$; mirror-type connector $[m_x\!\parallel\!\cspin]$. \\
\hline
(1/3,2/3) & AFM (Type II) & $P321$ (150) & 
$\{[h\!\parallel\!E]:h\in H\}\;\cup\;\spinlaue{P}{T}$ &
Effective $\pt$ via inversion pairing; no splitting. \\
\hline
(1/2,0) & AFM (Type II) & $Aem2$ (39) & 
$\{[h\!\parallel\!E]:h\in H\}\;\cup\;\spinlaue{P}{T}$ &
$2\mathbf{t}\in\mathcal{L}$ restores an effective $\pt$; spin degeneracy observed. \\
\bottomrule
\end{tabular}

\vspace{1em}
\textbf{AA$^\prime$ $\uparrow\downarrow/\downarrow\uparrow$ (Antiparallel) ordering} \\  
\begin{tabular}{@{} c | l | l | l | p{0.30\textwidth} @{}}
\toprule
Offset / Structure & Phase (Zeng et al.) & MSG & Spin–Laue Group & Generators / Comment \\
\midrule
(0,0) & AM (Type III / ${}^{1}\overline{3}^{2}m$) & $P\overline{6}'2m'$ (189.223) & 
$\{[h\!\parallel\!E]:h\in H\}\;\cup\;\spinlaue{C_2}{\cspin}$ &
Broken $\pt$; $[C_2\!\parallel\!\cspin]$ connector survives. \\
\hline
(0,1/3) & AM (Type III / ${}^{2}2/^{2}m_x$) & $C'm$ (8.32) & 
$\{[h\!\parallel\!E]:h\in H\}\;\cup\;\spinlaue{m_x}{\cspin}$ &
Broken $\pt$; mirror-type connector $[m_x\!\parallel\!\cspin]$. \\
\hline
(1/3,1/3) & AM (Type III / ${}^{2}2/^{2}m_x$) & $C'm$ (8.32) & 
$\{[h\!\parallel\!E]:h\in H\}\;\cup\;\spinlaue{m_x}{\cspin}$ &
Broken $\pt$; mirror-type connector $[m_x\!\parallel\!\cspin]$. \\
\hline
(1/3,2/3) & AM (Type III / ${}^{1}\overline{3}^{2}m$) & $P321'$ (150.25) & 
$\{[h\!\parallel\!E]:h\in H\}\;\cup\;\spinlaue{C_2}{\cspin}$ &
Broken $\pt$; $[C_2\!\parallel\!\cspin]$ connector present. \\
\hline
(1/2,0) & AM (Type III / ${}^{2}m^{2}m^{1}m$) & $A'e'm2$ (39.197) & 
$\{[h\!\parallel\!E]:h\in H\}\;\cup\;\spinlaue{m_x}{\cspin}$ &
Broken $\pt$; mirror-type connector with spin inversion yields Type III altermagnetism. \\
\bottomrule
\end{tabular}

\begin{flushleft}
\textit{Note:} Type II AFM phases possess spin degeneracy via $\mathcal{PT}$; Type III altermagnets require surviving non-degeneracy-enforcing connectors of the form $\spinlaue{a}{\cspin}$, with $a \notin \{P, M_z, C_{2z}\}$~\cite{Zeng2024Description,Smejkal2022}. The restoration of $\mathcal{PT}$ in the parallel $(1/2,0)$ case is described as “effective,” since it originates from the composite stacking operation $[P \parallel \tau_z]$ combined with time reversal~\cite{Pan2024,ramazashvili2009kramers}.
\end{flushleft}

\end{table*}

Stacking geometry and interlayer spin alignment provide two independent yet combinatorially effective symmetry control parameters whose interplay deterministically selects the magnetic phase in bilayer MnPS$_3$. In AA stacking, the parallel ($\uparrow\downarrow/\uparrow\downarrow$) order preserves $\pt$, yielding a robust Type II antiferromagnet. The antiparallel ($\uparrow\downarrow/\downarrow\uparrow$) order breaks it, opening a switchable pathway to either structured Type III altermagnetism when a non-degeneracy-enforcing connector $\SL{a}{\cspin}$ survives, or to a symmetry-starved Quasi-AM (SST-4) if no such connector remains. AA$^\prime$ stacking intrinsically breaks $\pt$ via layer inversion and therefore generically hosts Type III altermagnetism in the antiparallel case, with a controlled return to Type II AFM only at the special high-symmetry shift $(1/2,0)$ where an effective $\pt$ is recovered. The AA $(1/3,2/3)$, $\uparrow\downarrow/\downarrow\uparrow$ configuration provides a concrete SST-4 realization: lacking both $\pt$ and any $\SL{a}{\cspin}$ connector, it exhibits nonrelativistic spin splitting throughout the Brillouin zone, including at $\Gamma$, without the symmetry-enforced alternation of true altermagnets. This symmetry-engineered tunability establishes a practical design principle for realizing compensated magnetic phases in two-dimensional van der Waals materials.~\cite{Bai2025, Zeng2024Description, Pan2024, yuan2024nonrelativistic}

%

\noindent\textbf{Notes on Symmetry Tolerances}
\label{subsec:tolerances}

The identification and classification of magnetic phases in bilayer MnPS$_3$ hinge sensitively on small structural distortions and the numerical thresholds used in symmetry extraction. Idealized geometries might formally admit degeneracy-enforcing operations such as $[C_{2z}\Vert E]$ or $\mathcal{PT}$, but atomic relaxations at the 0.01--0.02~\AA\ scale can eliminate them once realistic tolerances are applied. These displacements are comparable to thermal root-mean-square atomic fluctuations at temperatures of order 100--300~K, as estimated from Debye–Waller physics where $\langle u^2\rangle^{1/2}\sim0.01$--0.03~\AA\ for typical van der Waals crystals~\cite{paris2016temperature,lin2017ultrafast}, underscoring that some apparent symmetry breaking is physically plausible even in well-relaxed structures.

To avoid misclassification, we choose tolerances that strike a balance between (i) retaining spurious symmetries that contradict electronic structure (false positives) and (ii) overpruning genuine, physically relevant symmetries (false negatives). Specifically, we adopt position tolerances up to 0.1~\AA\ and angular deviations up to $5^\circ$, and verify stability across reasonable tolerance variations to ensure that reported magnetic space groups and connector survivals are not artifacts of a particular threshold choice. When a candidate symmetry lies near the detection boundary or yields tension with other diagnostics, we do not accept it purely on algorithmic grounds. Instead, we corroborate its relevance by comparing to the DFT band structure: the presence or absence of nonrelativistic spin splitting, especially at high-symmetry points such as $\Gamma$, serves as a direct electronic fingerprint of whether a degeneracy-enforcing symmetry (e.g., $\mathcal{PT}$) truly survives or whether a purported connector is physically operative.
This dual approach, tolerance-aware group-theoretic extraction supplemented by first-principles spectral validation, ensures that the phase labels reflect the actual symmetry constraints experienced by the magnetic state, not numerical coincidences.

\section*{Discussion}

Our findings demonstrate that stacking configuration and magnetic alignment act as intertwined symmetry control parameters, enabling the deliberate switching of bilayer MnPS$_3$ between distinct magnetic phases. This symmetry-driven mechanism produces tunable momentum-dependent spin splitting in the electronic band structure, even in the absence of spin-orbit coupling, a defining characteristic of altermagnetic systems.
To quantify these effects, we extract spin splittings $\delta E(\mathbf{k}) = \epsilon_\uparrow(\mathbf{k}) - \epsilon_\downarrow(\mathbf{k})$ from the calculated band structures (Figures~\ref{fig:bandsAA} and~\ref{fig:bandsAAp}), averaging them over the Brillouin zone to obtain stacking-resolved trends. As summarized in Figure~\ref{fig:split}, particular stacking displacements, notably the AA$^\prime$ arrangement with $\uparrow\downarrow/\downarrow\uparrow$ spin order, yield momentum-dependent spin splittings reaching approximately 60--80 meV. These splittings, prominent near high-symmetry points such as $K$ and $\Gamma$, are comparable to those in other 2D altermagnets like MnTe~\cite{MnTe2024AMR} and induced monolayers~\cite{mazin2023induced}. Crucially, their magnitude exceeds the thermal energy scale at room temperature, suggesting robustness against thermal smearing and experimental visibility.
Such sizable splittings present clear experimental opportunities, especially via spin- and momentum-resolved angle-resolved photoemission spectroscopy (ARPES). The observed asymmetry between $K$ and $K'$ valleys implies emergent valley-contrasting spin textures, accessible through circular dichroism ARPES or nonlinear optics, such as second-harmonic generation. Specifically, we predict splittings of 60--80 meV in occupied bands near $\Gamma$ and $K$, within the resolution of modern ARPES systems.

Beyond spectroscopy, broken $\mathcal{P}\mathcal{T}$ symmetry and non-centrosymmetric stacking lead to anisotropic spin-split Fermi surfaces. This supports unconventional transport phenomena, including angle-dependent Hall and spin Hall drag~\cite{Lin2025CoulombDrag}, with predicted drag resistivities of 10--100~$\Omega$ and angular modulations governed by the layer's spin-Laue symmetry. Maximum drag is expected when current flows along high-symmetry directions, reflecting the momentum-dependent spin polarization.
Complementary transport experiments, such as angular-dependent magnetoresistance (AMR), offer another robust probe. Following demonstrations in MnTe~\cite{MnTe2024AMR}, we predict that reorienting the Néel vector in MnPS$_3$ bilayers will induce resistivity anisotropies, with twofold or sixfold modulations tied to stacking geometry and magnetic alignment.
Following the symmetry-based framework of Sheoran and Dev~\cite{sheoran2025spontaneous}, we expect AA$^\prime$ stacking with $\uparrow\downarrow/\downarrow\uparrow$ order to realize a $d$-wave altermagnetic phase that allows a symmetry-permitted in-plane anomalous Hall effect (AHE). In contrast, AA stacking preserves $\mathcal{P}\mathcal{T}$ symmetry, suppressing such responses. This enables detection of Néel vector orientation through transverse Hall signals, complementing AMR and drag measurements. Estimated AHE conductivities ($\sigma_{xy}$) in MnPS$_3$ bilayers, by analogy to MnPSe$_3$ and MnSe, range from 20 to 100 S/cm~\cite{sheoran2025spontaneous}, aligning with ARPES-predicted splittings near $K$ and $\Gamma$ of 60--80 meV, consistent with values in MnTe~\cite{MnTe2024AMR} and symmetry-induced systems~\cite{mazin2023induced}. For AMR, observed modulations of 3--5\% in MnTe suggest expected signals of 2--6\% in MnPS$_3$, depending on stacking and interlayer coupling.
Beyond these transport and spectroscopic signatures, recent theory suggests that 2D altermagnets may host unconventional superconducting phenomena when integrated into superconductor-magnet hybrid devices~\cite{fukaya2025superconducting}. In particular, $d$-wave altermagnets like bilayer MnPS$_3$ could enable spin-singlet to spin-triplet conversion, anomalous Josephson effects, and symmetry-driven superconducting diode behavior, even without net magnetization or spin-orbit coupling.

While our study focuses on MnPS$_3$, the underlying mechanisms are broadly applicable to other van der Waals bilayers with antiferromagnetic order. Systems such as NiPS$_3$ and CrI$_3$, with their distinct anisotropies, interlayer couplings, and spin-orbit strengths, offer promising platforms for stacking-tunable altermagnetic behavior. Moreover, the ability to control spin splitting in a layer-resolved fashion opens avenues for advanced multifunctional applications. In particular, integrating stacking-induced altermagnetism with intercalated molecules or external electric fields enables coupling to ferroelectric switching~\cite{Gu2025FEAltermagnetism}, laying the groundwork for non-volatile memory and tunable magnetoelectric devices that combine electric writing with magnetic reading.

Finally, the range of experimental signatures discussed, such as ARPES splittings, angle-dependent drag, anisotropic magnetoresistance, and ferroelectric coupling provides a robust framework for probing and harnessing stacking-tunable altermagnetism.

Density functional theory combined with spin–Laue symmetry analysis demonstrates that interlayer stacking geometry and relative spin alignment serve as effective symmetry-tuning knobs that determine the magnetic phase of bilayer MnPS$_3$. The monolayer is a conventional collinear antiferromagnet with no intrinsic momentum-dependent spin splitting. In the bilayer, particular combinations of stacking offset and interlayer order selectively break or preserve spin–space symmetries, enabling transitions among Type II antiferromagnetic, Type III altermagnetic, and quasi-altermagnetic (SST-4) regimes.\\

The essential distinction lies in the surviving spin–Laue group: if it retains a degeneracy-enforcing antiunitary symmetry such as $\pt$ (Type II), spin degeneracy persists; if $\pt$ is broken but a proper spin-exchanging connector $\SL{a}{\cspin}$ survives with $a\notin\langle \mathcal{P}, C_{2z}, M_z\rangle$, the system realizes Type III altermagnetism with alternating, momentum-dependent splitting. Pure spatial operations like $\SL{C_{2z}}{E}$ or their nonsymmorphic relatives $\SL{C_{2z}}{\mathbf{t}}$ do not induce splitting unless accompanied by the spin inversion $\cspin$. When neither $\pt$ nor any valid $\SL{a}{\cspin}$ remains, the phase enters the quasi-altermagnetic (SST-4 / Type I) regime, where splitting reflects symmetry starvation rather than a protected pattern. Notably, $\pt$, despite $(\mathcal{P}\mathcal{T})^2=+1$, enforces a Kramers-like degeneracy in compensated antiferromagnets, stabilizing the Type II phase~\cite{Bai2025}.

In AA stacking, the $\uparrow\downarrow/\uparrow\downarrow$ order universally preserves $\pt$, producing a symmetry-protected Type II antiferromagnet across lateral shifts. The $\uparrow\downarrow/\downarrow\uparrow$ alignment breaks $\pt$, and the emergent phase depends on connector survival: if a non-degeneracy-enforcing connector such as $\SL{C_{2z}}{\cspin}$, $\SL{C_{2,[110]}}{\cspin}$, or $\SL{m_x}{\cspin}$ survives, the system shows Type III altermagnetic splitting; if no such connector remains (notably at $\mathbf{t}=(1/3,2/3)$), the result is a Quasi-Altermagnet, a concrete SST-4 realization that lacks both $\pt$ and any valid $\SL{a}{\cspin}$ yet exhibits nonrelativistic spin splitting even at $\Gamma$ without the symmetry-enforced alternation of true altermagnets.

In AA$^\prime$ stacking, the inversion-related symmetries become more nuanced due to the internal layer inversion. For antiparallel order, $\pt$ is broken in all cases, but surviving non-degeneracy-enforcing connectors consistently link opposite-spin sublattices, making Type III altermagnetism the generic outcome. For parallel order, an effective $\pt$ can re-emerge only at special high-symmetry offsets satisfying $2\mathbf{t}\in\mathcal{L}$ via the composite stacking-operator mechanism of Pan et al.~\cite{Pan2024}, restoring spin degeneracy and reverting the system to a Type II AFM.\\

Our DFT results corroborate this symmetry-based classification, distinguishing (i) phases with exact $\pt$ that exhibit no intrinsic nonrelativistic splitting (Type II AFM), (ii) phases where opposite-spin sublattices are connected by a valid $\SL{a}{\cspin}$, producing the momentum-dependent splitting of Type III altermagnets, and (iii) quasi-altermagnetic states lacking both protections, where splitting reflects the absence of symmetry constraints rather than a symmetry-generated pattern. The mechanism stacking-controlled emergence or suppression of spin-exchanging symmetry operations, is not unique to MnPS$_3$ and should extend to other van der Waals antiferromagnets with tunable interlayer registry. The resulting modifications of the band structure, particularly the presence, symmetry, and orientation of nonrelativistic spin splitting near high-symmetry points, are accessible to spin- and momentum-resolved probes such as spin-resolved ARPES.

\section*{Methods}

\noindent\textbf{Computational Details}

We carried out density functional theory (DFT) calculations using the Perdew-Burke-Ernzerhof (PBE) generalized gradient approximation~\cite{PBE}, as implemented in the Vienna \textit{ab initio} Simulation Package (VASP)~\cite{VASP}, within the plane-wave pseudopotential framework. To account for London dispersion forces, we employed the optB86b-vdW functional~\cite{optB86b}. A plane-wave energy cutoff of 500 eV was used. Monkhorst-Pack $8\times8\times1$ and $12\times12\times1$ $k$-point meshes were adopted for structural relaxation and self-consistent calculations, respectively~\cite{Monkorst}. Electronic correlations in Mn atoms were treated using a Hubbard on-site Coulomb parameter of $U = 5$ eV within the Dudarev approach~\cite{dudarev1998electron}. \bluemark{This value is well established in the literature for MnPS$_3$, as it reproduces key experimental observables such as the band gap and magnetic moment\cite{yang2020electronic,autieri2022limited}. For all stacking configurations and magnetic alignments studied, the total magnetic moment per unit cell remains zero within $10^{-4},\mu_B$, confirming that each phase is fully compensated and satisfies the defining zero-net-magnetization condition for altermagnetism.}

\noindent\textbf{Constrained Relaxation}

To preserve the intended stacking during relaxation, we fixed the in-plane ($x,y$) positions of the outermost atoms (including Mn) while allowing relaxation along $z$, with all inner atoms fully relaxed. This ensures optimization of interlayer distances and bonding while maintaining the target geometry. Convergence was reached when residual forces on unconstrained atoms were below $10^{-3}$ eV/Å. While such constraints are required to construct the energy map~\cite{sivadas2018stacking,cortes2018stacking,leon2025interlayer}, we also performed fully unconstrained relaxations, which confirmed that the high-symmetry stacking configurations $(m/3, n/3)$ with $m,n = 0,1,2$ are indeed stable.

\section*{Data availability}
All data supporting the findings of this study are included in the article and its Supplementary Information.

\section*{Code Availability}
Density functional theory calculations were performed using the Vienna \textit{ab initio} Simulation Package (VASP), accessed under an institutional license through the high-performance computer at the NHR Center of TU Dresden and on the high-performance computers Noctua 2 at the NHR Center PC2.  
Custom Python scripts developed for post-processing and analysis are available from the corresponding author upon reasonable request.

\section*{Acknowledgments}
J.W.G. acknowledges financial support from ANID-FONDECY Grant No.~1220700 (Chile). 
E.S.M. and J.W.G. also acknowledge support from ANID-FONDECYT Grant No.~1221301 (Chile).  
T.B. thanks the German Research Foundation (DFG) for funding through the DIP Grant HE 3543/42-1 (Project No.~471289011).  
The authors gratefully acknowledge computing time provided by the NHR Center at TU Dresden and the Noctua 2 system at the NHR Center PC2. These resources are funded by the German Federal Ministry of Education and Research and the participating state governments, based on resolutions of the GWK for national high-performance computing at universities (\url{www.nhr-verein.de/unsere-partner}).  
A.L. gratefully acknowledges support from the Technische Universität Dresden Professor Fellowship Program. This research was initiated during her visit to the Institute for Solid State and Materials Physics in Dresden.\\

\section*{Author Contributions}
A.L. and J.W.G. performed the calculations and wrote the initial draft. 
All authors contributed to the discussion of the results and collaborated in editing and shaping the final version of the manuscript.

\section*{Competing interests}
The authors declare no competing interests.

\end{document}